\documentclass[12pt,twoside]{article}

\usepackage{epsf,amssymb}

\begin{document}

\thispagestyle{empty}
\vspace*{2.5cm}
\begin{flushleft}
\hspace{10mm}
\uppercase{Classical properties of generalized coherent}\\ 
\hspace{10mm}
\uppercase{states: from phase-space dynamics to Bell's}\\ 
\hspace{10mm} 
\uppercase{inequality} \\
\vspace{3\baselineskip}
\hspace{10mm}
C. BRIF, \underline{A. MANN}, and M. REVZEN \\
\vspace{1\baselineskip}
\hspace{10mm}
Department of Physics, Technion---Israel Institute of
Technology, \\
\hspace{10mm} Haifa 32000, Israel
\end{flushleft}
\vspace{1\baselineskip}

\noindent
\hspace{10mm}
\parbox[b]{130mm}{{\bf Abstract}\hspace{0.4cm}
We review classical properties of harmonic-oscillator coherent 
states. Then we discuss which of these classical properties are 
preserved under the group-theoretic generalization of coherent 
states. We prove that the generalized coherent states of quantum 
systems with Lie-group symmetries are the unique Bell states, 
i.e., the pure quantum states preserving the fundamental 
classical property of satisfying Bell's inequality upon 
splitting.}
\vspace{1\baselineskip}

\noindent
\uppercase{1. Introduction}
\vspace{1\baselineskip}

\noindent
A few years ago, a joint paper with Prof.\ Ezawa$^{1}$ started:
``Some 25 years ago Aharonov, Falkoff, Lerner, and Pendleton
characterized the quantum state of the radiation field known
as coherent state by a classical attribute it possesses.''
In the present work dedicated to Prof.\ Ezawa upon his 65th
birthday (Dear Hiroshi: Tanjobi omedeto gozaimas!) we would 
like to discuss a few more aspects of the quantum-classical
relationship.
\vspace{1.6\baselineskip}

\noindent
\uppercase{2. Harmonic-oscillator coherent states}
\vspace{0.8\baselineskip}

\noindent
Coherent states (CS) for a quantum harmonic oscillator were first
discovered in 1926 by Schr\"{o}dinger$^{2}$ who looked for wave 
packets with minimum possible dispersions that will preserve 
their form while moving along a classical trajectory.
An enormous interest in the CS was stimulated by pioneering works 
of Glauber$^{3}$ who introduced them in the context of quantum 
optics and invented their name.
Glauber proposed two criteria for the CS:
(i) quantum states of the radiation field produced by a 
classically prescribed current; 
(ii) states possessing the property of maximal coherence (which is 
the origin of the name): normal-ordered correlation functions of 
all orders factorize.

There are two (equivalent) formal definitions of the CS:
(i) eigenstates of the boson annihilation operator:
\begin{equation}
a |\alpha\rangle = \alpha |\alpha\rangle , 
\hspace{10mm} \alpha \in \mathbb{C} 
\label{def1} 
\end{equation}
(this property actually implies another possible definition of 
$|\alpha\rangle$ as minimum-uncertainty states);
(ii) states produced by the action of the displacement 
operator $D(\alpha)$ on the vacuum,
\begin{equation}
|\alpha\rangle = D(\alpha) |0\rangle = 
\exp(\alpha a^{\dagger} - \alpha^{\ast} a) |0\rangle .
\label{def2} 
\end{equation}

There exists an intimate relation between the CS and the concept
of phase space in quantum mechanics.
The scaled position and momentum of a quantum harmonic oscillator
(which is the mathematical model of a single mode of the quantized
radiation field) are given by
\begin{equation}
q = \frac{1}{\sqrt{2}} (a+a^{\dagger}) , \hspace{10mm}
p = \frac{1}{ i \sqrt{2}} (a-a^{\dagger}) .
\end{equation}
For the CS $|\alpha\rangle$ one obtains:
\begin{equation}
\langle q \rangle = \sqrt{2}\, \mathrm{Re}\, \alpha , \hspace{8mm}
\langle p \rangle = \sqrt{2}\, \mathrm{Im}\, \alpha , \hspace{8mm}
\Delta q = \Delta p = 1/\sqrt{2} . 
\end{equation}
Therefore the phase space $(q,p)$ is just the complex $\alpha$ 
plane. 

The coherent-state wave packets not only have minimal possible
dispersions, but also preserve this property during the
dynamical evolution. 
The evolution of a single-mode quantized radiation field 
interacting with a classical external source is governed by 
the Hamiltonian
\begin{equation}
H = \omega a^{\dagger} a + \lambda(t) a^{\dagger} 
+ \lambda^{\ast}(t) a ,
\end{equation}
which is linear in the generators of the oscillator
group $H_4$.
If the initial state $|\psi(0)\rangle$ is the vacuum, the solution 
of the Schr\"{o}dinger equation is$^4$
\begin{eqnarray}
|\psi(t)\rangle & = & {\cal T} 
\exp\left( - i \int H(t) d t \right) |0\rangle \nonumber \\
& = & e^{ i \eta(t) } \exp\left[ \alpha(t) a^{\dagger} 
- \alpha^{\ast}(t) a \right] |0\rangle
= e^{ i \eta(t) } |\alpha(t)\rangle ,
\end{eqnarray}
where
\begin{eqnarray*}
& & \alpha(t) = - i e^{- i \omega t} \int_{0}^{t}
\lambda^{\ast}(\tau) e^{ i \omega \tau } d \tau , \\
& & \eta(t) = -\frac{\omega t}{2} -  \int_{0}^{t}
\mathrm{Re}[\lambda(\tau) \alpha(\tau)] d \tau .
\end{eqnarray*}
This result means that the system will remain for all times 
in a coherent state. Furthermore, if the initial state is a
coherent state (including the vacuum) and the Hamiltonian
is linear in the generators of $H_4$, then the state will evolve
into a coherent state, i.e., ``{\em once a coherent state, 
always a coherent state}''. 
The CS will evolve along a classical trajectory 
\begin{equation}
\alpha(t) = \left[ q(t) + i p(t) \right]/\sqrt{2}
\end{equation}
in the phase space. In the next section we will consider
how this important property of the harmonic-oscillator CS
can be generalized for other quantum systems.
\vspace{1.6\baselineskip}

\noindent
\uppercase{3. The generalized coherent states}
\vspace{0.8\baselineskip}

\noindent
The Gilmore-Perelomov generalization$^{4,5}$ of the CS
consists of the following components:
$G$ is a Lie group (the dynamical symmetry group of a quantum 
system), $\Gamma_{\Lambda}$ is unitary irreducible 
representation (irrep) of $G$ acting on 
the Hilbert space ${\cal H}_{\Lambda}$, $|\Psi_{0}\rangle$ is 
a fixed normalized reference state in ${\cal H}_{\Lambda}$.
The generalized CS $|\Psi_{g}\rangle$ are produced by the 
action of group elements on the reference state:
\begin{equation}
|\Psi_{g}\rangle = \Gamma_{\Lambda}(g) |\Psi_{0}\rangle , 
\hspace{1cm} g \in G . 
\end{equation}
Elements of the isotropy subgroup $H \subset G$ leave the 
reference state invariant up to a phase factor:
\begin{equation}
\Gamma_{\Lambda}(h) |\Psi_{0}\rangle = 
e^{i\phi(h)} |\Psi_{0}\rangle , 
\hspace{1cm} h \in H .               
\end{equation}
For every element $g \in G$, there is a decomposition:
\begin{equation}
g = \Omega h , \hspace{1cm} g \in G, \;\; h \in H, \;\; 
\Omega \in G/H     
\end{equation}
where $G/H$ is the coset space. 
Group elements $g$ and $g'$ with different $h$ and $h'$ but with 
the same $\Omega$ produce CS which differ only by a phase factor: 
$|\Psi_{g}\rangle = e^{i\delta} |\Psi_{g'}\rangle$, where 
$\delta =\phi(h) -\phi(h')$.
Therefore a coherent state 
$|\Omega\rangle \equiv |\Psi_{\Omega}\rangle$ 
is determined by a point $\Omega = \Omega(g)$ 
in the coset space.

The standard (maximum-symmetry) system of the CS is obtained when 
the reference state is an `extreme' state of the Hilbert space 
(e.g., the vacuum state of an oscillator or the lowest/highest 
spin state). Then $G/H$ will be a homogeneous K\"{a}hlerian 
manifold. 
{\em Then the coset space is the phase space of the system.} 
Examples of phase spaces:
(1) The Glauber CS of the Heisenberg-Weyl group H$_{3}$ are 
defined on the complex plane 
$\mathbb{C} = \mathrm{H}_3 / \mathrm{U}(1)$.
(2) The spin CS are defined on the unit sphere 
$\mathbb{S}^2 = \mathrm{SU}(2) / \mathrm{U}(1)$.

Under the action of the group elements the generalized CS
transform among themselves. If the Hamiltonian is linear in the 
group generators
\begin{equation}
H = \sum_{i} \beta_i T_i ,
\end{equation}
the evolution operator will be an element of the group.
{\em Then the generalized CS will evolve along a classical 
trajectory $\Omega(t)$ in the phase space.}
\vspace{1.6\baselineskip}

\noindent
\uppercase{4. Another criterion of classicality}
\vspace{0.8\baselineskip}

\noindent
In 1966 Aharonov, Falkoff, Lerner and Pendleton (AFLP) 
proposed$^6$ a characterization of a quantum state of the radiation 
field by a classical attribute it possesses. They used the
following arguments: In classical physics, two observers $B$ and $C$ 
cannot ascertain by any local measurements (including correlating 
their observations) whether two radiation beams emanated from one 
source that was split---or they came from two independent sources 
(see Figure 1). In quantum physics, quantum correlations created 
between the beams upon splitting will, in general, enable the 
observers to distinguish between the two situations.

\vspace{3mm}
\begin{figure}[htbp]
\epsfxsize=0.6\textwidth
\centerline{
\epsffile{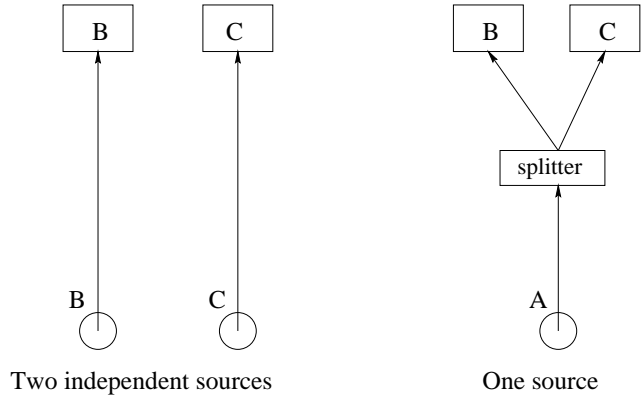}}
\vspace*{1mm}
\caption{A sketch of the AFLP scheme.}
\end{figure}

However, the Glauber CS are the only quantum states of the
radiation field which possess the classical attribute:
The beams $B$ and $C$ that were split from one source $A$
can be simulated by two beams from independent sources.
The reason is that no quantum correlations are created for 
the coherent state upon splitting.
This means that the field coherent state will factorize upon
splitting and the beams $B$ and $C$ will be disentangled.

We present here a brief description of the proof given by 
AFLP$^6$. A state of the radiation field in the mode $A$ can
be written as 
\begin{equation}
|\psi_{A}\rangle_{A} = f_{A}(a_{A}^{\dagger}) |0\rangle , 
\end{equation}
and analogously for the modes $B$ and $C$.
Splitting of the light beam can be realized by a half-silvered 
mirror (with the vacuum in the second input port).
The corresponding transformation is
\begin{equation}
a_{A}^{\dagger} = \mu a_{B}^{\dagger} + \nu a_{C}^{\dagger},
\hspace{10mm} |\mu|^2 + |\nu|^2 = 1 . 
\end{equation}
One looks for the state that factorizes upon splitting,
\begin{equation}
|\psi_{A}\rangle_{A} = 
|\psi_{B}\rangle_{B} \otimes |\psi_{C}\rangle_{C} .   
\end{equation}
This leads to the functional equation:
\begin{equation}
f_{A}(\mu a_{B}^{\dagger} + \nu a_{C}^{\dagger}) 
= f_{B}(a_{B}^{\dagger}) f_{C}(a_{C}^{\dagger}) . 
\end{equation}
Following AFLP, the unique solution of this equation
is given by the functions (including normalization):
\begin{eqnarray*}
& & f_{A}(a_{A}^{\dagger}) = D_{A}(\alpha) =
\exp(\alpha a_{A}^{\dagger} - \alpha^{\ast} a_{A}), \\
& & f_{B}(a_{B}^{\dagger}) = D_{B}(\alpha_{B}) =
\exp(\alpha_{B} a_{B}^{\dagger} - \alpha^{\ast}_{B} a_{B}), \\
& & f_{C}(a_{C}^{\dagger}) = D_{C}(\alpha_{C}) =
\exp(\alpha_{C} a_{C}^{\dagger} - \alpha^{\ast}_{C} a_{C}), 
\end{eqnarray*}
where 
$\alpha_{B} = \mu \alpha_{A}$, $\alpha_{C} = \nu \alpha_{A}$.
These operators are exactly the displacement operators
producing the Glauber CS. Therefore, the Glauber CS are
the unique field states which factorize upon splitting:
\begin{equation}
|\alpha\rangle_A = |\mu\alpha\rangle_B \otimes
|\nu\alpha\rangle_C , \hspace{10mm} |\mu|^2 + |\nu|^2 = 1 .
\label{aflpcase}
\end{equation}
\vspace{0.1\baselineskip}

\noindent
\uppercase{5. Bell's inequality and Bell states}
\vspace{0.8\baselineskip}

\noindent
A state of a quantum system, consisting of two subsystems, is
called {\em entangled}, if it cannot be represented as a direct 
product of states of the subsystems.
This situation is in contradiction with classical physics as 
expressed mathematically by the violation of Bell's 
inequality$^7$.

Let $B$ and $C$ be the two subsystems. The operator 
$\hat{C}(\rho)$ acts on the subsystem $C$ and has eigenvalues
$\pm 1$, depending on the parameter $\rho$. The operator 
$\hat{B}(\sigma)$ acts in the same way on the subsystem $B$.
Then Bell's inequality (in the version of Clauser, Horne,
Shimony, and Holt$^8$) reads
\begin{equation}
\left| \langle \psi | \hat{C}(\rho) \hat{B}(\sigma) +
\hat{C}(\rho) \hat{B}(\sigma') 
+ \hat{C}(\rho') \hat{B}(\sigma) - \hat{C}(\rho') 
\hat{B}(\sigma') | \psi \rangle \right| \leq 2 .   
\label{bellineq}
\end{equation}
It is clear that the product state
$$ |\psi\rangle = |\psi_{B}\rangle_{B} \otimes 
|\psi_{C}\rangle_{C} $$
satisfies Bell's inequality (\ref{bellineq}).
Furthermore, it has been proven$^{9-11}$ that any entangled 
(i.e., non-product) pure state will always violate the 
inequality. In other words, for pure states there is 
equivalence between entanglement and the violation of Bell's 
inequality.

Let us now consider a quantum system $A$ which is split or 
decays into two subsystems $B$ and $C$. 
A pure quantum state that upon splitting will not violate 
Bell's inequality is called the {\em Bell state}$^{11}$. 
This definition means that the Bell state of the system $A$ 
must factorize upon splitting into the direct product of 
states of the subsystems $B$ and $C$. 
Then the AFLP result means that the Glauber CS are the unique
Bell states of the quantized radiation field.

Let us address the following question: What are the Bell 
states for quantum systems other than the single-mode radiation 
field? Does this classical attribute exist for a quantum 
system of general symmetry?
\vspace{1.6\baselineskip}

\noindent
\uppercase{6. Coherent states for semisimple Lie groups}
\vspace{0.8\baselineskip}

\noindent
We will consider quantum systems, whose dynamical symmetry
groups are semisimple Lie groups. In particular, we will
consider SU(2) which is the dynamical symmetry
group of a spin-$j$ particle. We use the following notation:
$G$ is a semisimple Lie group,
$\mathfrak{G}$ is the corresponding Lie algebra,
$\mathfrak{H}$ is the Cartan subalgebra,
$\Delta$ is the set of non-zero roots,
$\{ H_{i},E_{\alpha} \}$ is the Cartan-Weyl basis
($H_{i} \in \mathfrak{H}$; $\alpha,\beta \in \Delta$):
\begin{eqnarray*}
& & [H_{i},H_{j}] = 0 , \\
& & [H_{i},E_{\alpha}] = \alpha(H_{i}) E_{\alpha} , \\
& & [E_{\alpha},E_{\beta}] = \left\{ 
  \begin{array}{l}
0 , \;\;\;\; {\rm if}\; \alpha + \beta \neq 0\; {\rm and}\;
\alpha + \beta \notin \Delta , \\
H_{\alpha} , \;\;\;\; {\rm if}\; \alpha + \beta = 0 , \\
N_{\alpha,\beta} E_{\alpha + \beta} , \;\;\;\; {\rm if}\;
\alpha + \beta \in \Delta .
  \end{array} \right.
\end{eqnarray*}
The generators $H_i$ are diagonal in any irrep $\Gamma_{\Lambda}$,
while $E_{\alpha}$ are the ``shift operators''. For a Hermitian 
irrep, one has $H_{i}^{\dagger} = H_{i}$ and 
$E_{\alpha}^{\dagger} = E_{-\alpha}$.

A physically sensible choice of the reference state 
$|\Psi_{0}\rangle$ is the ground state of the system, i.e.,
an ``extremal state'' in the Hilbert space ${\cal H}_{\Lambda}$.
{\em This choice of the reference state determines the 
classical properties of the CS.}
The extremal state is the lowest-weight state 
$|\Lambda,-\Lambda\rangle$:
\begin{eqnarray}
& & E_{\alpha} |\Lambda,-\Lambda\rangle = 0 , \hspace{10mm} 
\alpha < 0 , \\
& & H_{i} |\Lambda,-\Lambda\rangle = \Lambda_{i} 
|\Lambda,-\Lambda\rangle , 
\hspace{10mm} H_{i} \in \mathfrak{H} .
\end{eqnarray}
Thus the Lie algebra of the isotropy subgroup $H$ is just the 
Cartan subalgebra $\mathfrak{H}$.
Therefore, elements $\Omega$ of the coset space $G/H$ are 
\begin{equation}
\Omega = \exp\left[ \sum_{\alpha>0} ( \eta_{\alpha} E_{\alpha} 
- \eta_{\alpha}^{\ast} E_{-\alpha} ) \right] ,
\end{equation}
where $\eta_{\alpha}$ are complex numbers.
Using the Baker-Campbell-Hausdorff formula, one obtains
$$
\exp\!\left[ \sum_{\alpha>0} ( \eta_{\alpha} E_{\alpha} 
- \eta_{\alpha}^{\ast} E_{-\alpha} ) \right]\!  
= \exp\!\left[ \sum_{\alpha>0} \tau_{\alpha} E_{\alpha} \right]
\exp\!\left[ \sum_{i} \gamma_{i} H_{i} \right]
\exp\!\left[ -\sum_{\alpha>0} \tau_{\alpha}^{\ast} 
E_{-\alpha} \right]\! . 
$$
The relation between $\tau_{\alpha}$, $\gamma_{i}$ and 
$\eta_{\alpha}$ can be derived$^{4}$ from the matrix 
representation of $G$.
The generalized CS $|\Lambda,\Omega\rangle$ are given by
\begin{equation}
|\Lambda,\Omega\rangle = \Omega |\Lambda,-\Lambda\rangle
= {\cal N} \exp\left[ \sum_{\alpha>0} \tau_{\alpha} 
E_{\alpha} \right] |\Lambda,-\Lambda\rangle ,
\end{equation}
where the normalization factor is
\begin{equation}
{\cal N} = \exp\left[ \sum_{i} \gamma_{i} \Lambda_{i} \right] .
\label{nfactor}
\end{equation}

As an example we consider the SU(2) group for a spin-$j$ particle. 
The su(2) algebra is spanned by $\{ J_{0} , J_{+} , J_{-} \}$,
\begin{equation}
[J_{0} , J_{\pm}] = \pm J_{\pm}, \hspace{1.0cm} 
[J_{+} , J_{-}] = 2 J_{0} \ .   
\end{equation}
The Casimir operator
$J^{2} = J_{0}^{2} + (J_{+}J_{-} + J_{-}J_{+})/2$
for any irrep $\Gamma_{j}$ is the identity operator 
times a number: $J^{2} = j(j+1) I$ (where $j=0,1/2,1,3/2,\ldots$).
The lowest-weight state is $|j,-j\rangle$ (annihilated by $J_{-}$).
The isotropy subgroup $H$=U(1) consists of all group elements 
$h$ of the form $h=\exp(i\delta J_{3})$. 
The coset space is SU(2)/U(1) (the sphere), and 
the coherent state is specified by a unit vector
\begin{equation}
\mathbf{n} = (\sin\theta\cos\varphi,\sin\theta\sin\varphi,\cos\theta).
\end{equation}
Then an element $\Omega$ of the coset space is
\begin{equation}
\Omega = \exp (\xi J_{+} - \xi^{\ast} J_{-}) ,  
\end{equation}
where $\xi = -(\theta/2) e^{-i\varphi}$. 
The SU(2) CS are given by 
\begin{equation}
|j,\zeta\rangle =  \Omega |j,-j\rangle
= (1+|\zeta|^{2})^{-j} \exp(\zeta J_{+}) |j,-j\rangle , 
\end{equation}
where $\zeta = (\xi/|\xi|)\tan |\xi| = - \tan (\theta/2) 
e^{-i\varphi}$.
\vspace{1.6\baselineskip}

\noindent
\uppercase{7. Factorization upon splitting or decay}
\vspace{0.8\baselineskip}

\noindent
Consider a quantum system $A$ (whose dynamical symmetry group 
is $G$), that is split or decays into two subsystems $B$ 
and $C$ {\em of the same symmetry}.
The commutation relations are satisfied by operators
describing each of the three systems if and only if
\begin{eqnarray}
& & E_{A \alpha} = E_{B \alpha} \otimes I_{C} + 
I_{B} \otimes E_{C \alpha} ,  \label{decay:a} \\
& & H_{A i} = H_{B i} \otimes I_{C} + I_{B} \otimes H_{C i} ,
\label{decay:b}
\end{eqnarray}
where $\alpha \in \Delta$ and $H_{X i} \in \mathfrak{H}$, 
$X = A,B,C$.
For SU(2), this condition is equivalent to the rule for the 
addition of angular momenta,
\begin{equation}
\mathbf{J}_{A} =  \mathbf{J}_{B} \otimes I_{C} + 
I_{B} \otimes \mathbf{J}_{C} .
\end{equation}

We look for quantum states factorizing upon splitting.
First, we require that the lowest-weight state factorize:
\begin{equation}
|\Lambda_{A},-\Lambda_{A}\rangle_{A} =  
|\Lambda_{B},-\Lambda_{B}\rangle_{B} \otimes 
|\Lambda_{C},-\Lambda_{C}\rangle_{C} ,
\label{lws}
\end{equation}
giving the necessary condition for the factorization:
\begin{equation}
\Lambda_{A i} = \Lambda_{B i} + \Lambda_{C i} .
\label{neccond}
\end{equation}
In particular, for SU(2) this condition reads
\begin{equation}
j_{A} = j_{B} + j_{C} .
\label{jcond}
\end{equation}
If the representations of the subsystems $B$ and $C$ satisfy 
the condition (\ref{neccond}), then 
{\em any generalized coherent state factorizes upon splitting.}
Indeed, using Equations (\ref{decay:a}) and (\ref{lws}), we obtain
\begin{eqnarray*}
& & |\Lambda_{A},\Omega(\eta)\rangle_{A} = 
\exp\left[ \sum_{\alpha>0} ( \eta_{\alpha} E_{A \alpha} 
- \eta_{\alpha}^{\ast} E_{A \alpha}^{\dagger} ) \right]
|\Lambda_{A},-\Lambda_{A}\rangle_{A} \\
& & = \exp\left[ \sum_{\alpha>0} ( \eta_{\alpha} E_{B \alpha} 
- \eta_{\alpha}^{\ast} E_{B \alpha}^{\dagger} ) \right]
|\Lambda_{B},-\Lambda_{B}\rangle_{B} \\
& & \otimes \exp\left[ \sum_{\alpha>0} ( \eta_{\alpha} E_{C \alpha} 
- \eta_{\alpha}^{\ast} E_{C \alpha}^{\dagger} ) \right]
|\Lambda_{C},-\Lambda_{C}\rangle_{C} ,
\end{eqnarray*}
which gives
\begin{equation}
|\Lambda_{A},\Omega(\eta)\rangle_{A} = 
|\Lambda_{B},\Omega(\eta)\rangle_{B} 
\otimes |\Lambda_{C},\Omega(\eta)\rangle_{C} .
\end{equation}

The amplitudes $\eta_{\alpha}$ are the
same for the system $A$ and for the subsystems $B$ and $C$.
In particular, we find for spin
\begin{equation}
|j_{A},\zeta\rangle_{A} = 
|j_{B},\zeta\rangle_{B} \otimes |j_{C},\zeta\rangle_{C} ,
\end{equation}
i.e., $\zeta_{A} = \zeta_{B} = \zeta_{C}$. (The lowest and 
highest spin states $|j,-j\rangle$ and $|j,j\rangle$ are
particular cases of the SU(2) CS.)
The situation is somewhat different for the Glauber 
CS $|\alpha\rangle$ of the radiation field,
where the coherent-state amplitude is split according to
Equation (\ref{aflpcase}). 
The reason for this distinction lies in the differing structures
of the nilpotent group $H_{3}$ and a semisimple group $G$.

As a simple example we consider a spin-one particle 
(the system $A$) decaying into two spin-half particles 
(the subsystems $B$ and $C$). 
In this case
\begin{eqnarray*}
& & \hspace{-3mm} |1,1\rangle_{A} = 
|\mbox{$\frac{1}{2}$},\mbox{$\frac{1}{2}$}\rangle_{B} \otimes
|\mbox{$\frac{1}{2}$},\mbox{$\frac{1}{2}$}\rangle_{C} , 
\label{CGrel1} \\
& & \hspace{-3mm}
|1,0\rangle_{A} = \frac{1}{\sqrt{2}} \left( 
|\mbox{$\frac{1}{2}$},\mbox{$\frac{1}{2}$}\rangle_{B} \otimes
|\mbox{$\frac{1}{2}$},-\mbox{$\frac{1}{2}$}\rangle_{C} +
|\mbox{$\frac{1}{2}$},-\mbox{$\frac{1}{2}$}\rangle_{B} \otimes
|\mbox{$\frac{1}{2}$},\mbox{$\frac{1}{2}$}\rangle_{C} 
\right) ,
\label{CGrel2} \\
& & \hspace{-3mm} |1,-1\rangle_{A} = 
|\mbox{$\frac{1}{2}$},-\mbox{$\frac{1}{2}$}\rangle_{B} \otimes
|\mbox{$\frac{1}{2}$},-\mbox{$\frac{1}{2}$}\rangle_{C} . 
\label{CGrel3}
\end{eqnarray*}
The explicit form of the SU(2) CS is
\begin{equation}
|\mbox{$\frac{1}{2}$},\zeta \rangle = 
\frac{ |\mbox{$\frac{1}{2}$},-\mbox{$\frac{1}{2}$}\rangle  
+ \zeta |\mbox{$\frac{1}{2}$},\mbox{$\frac{1}{2}$}\rangle }{ 
\sqrt{1+|\zeta|^{2}} }
\label{jhcs}
\end{equation}
for $j=1/2$ and
\begin{equation}
|1,\zeta \rangle = \frac{ |1,-1\rangle + \sqrt{2} \zeta |1,0\rangle 
+ \zeta^{2} |1,1\rangle }{ 1+|\zeta|^{2} } 
\label{j1cs}
\end{equation}
for $j=1$.
Assume that the spin-one particle $A$ was prepared 
before the decay in the coherent state. Then we obtain
\begin{eqnarray}
|1,\zeta \rangle_{A} & = & \frac{  
|\mbox{$\frac{1}{2}$},-\mbox{$\frac{1}{2}$}\rangle_{B}  
+ \zeta |\mbox{$\frac{1}{2}$},\mbox{$\frac{1}{2}$}\rangle_{B} }{
\sqrt{1+|\zeta|^{2}} } \otimes 
\frac{ |\mbox{$\frac{1}{2}$},-\mbox{$\frac{1}{2}$}\rangle_{C}  
+ \zeta |\mbox{$\frac{1}{2}$},\mbox{$\frac{1}{2}$}\rangle_{C} }{ 
\sqrt{1+|\zeta|^{2}} } \nonumber \\
& = & |\mbox{$\frac{1}{2}$},\zeta \rangle_{B} \otimes
|\mbox{$\frac{1}{2}$},\zeta \rangle_{C} .
\label{csprod1}
\end{eqnarray}
The direct products $|1,1 \rangle_{A}$ and $|1,-1 \rangle_{A}$ 
are obtained as particular cases of (\ref{csprod1}) for 
$\zeta \rightarrow \infty$ and $\zeta = 0$, respectively.

We can prove$^{12}$ that the generalized CS are the {\em only\/} 
states which factorize upon splitting.
We first note that the states of the orthonormal basis are 
obtained by applying the raising operators to the lowest-weight 
state one or more times:
\begin{equation}
(E_{\alpha})^p |\Lambda,-\Lambda\rangle = 
|\Lambda,-\Lambda+p\alpha\rangle \times {\rm factor} ,
\end{equation}
where $\alpha > 0$ and $p \in \mathbb{N}$.
Therefore, any state $|\Phi\rangle \in {\cal H}_{\Lambda}$
can be obtained by applying a function of the raising operators
to the lowest-weight state:
\begin{equation}
|\Phi\rangle = f(\{E_{\alpha}\}) |\Lambda,-\Lambda\rangle , 
\hspace{1cm} \alpha > 0 .
\end{equation}
For example, for the SU(2) group one has
\begin{equation}
|j,m\rangle = \left( \begin{array}{c} 2j \\ j+m \end{array}
\right)^{1/2} \frac{ (J_{+})^{j+m} }{ (j+m)! } |j,-j\rangle ,
\end{equation}
and $|\Phi\rangle = f(J_{+}) |j,-j\rangle$.

If the state $|\Phi_{A}\rangle_{A}$ factorizes 
upon splitting, 
\begin{equation}
|\Phi_{A}\rangle_{A} = |\Phi_{B}\rangle_{B} \otimes
|\Phi_{C}\rangle_{C} ,
\end{equation}
then the following functional equation must be satisfied,
\begin{equation}
f_{A}(\{E_{A \alpha}\}) = f_{B}(\{E_{B \alpha}\})
f_{C}(\{E_{C \alpha}\})   
\end{equation}
(here and in what follows $\alpha > 0$).
Since $E_{A \alpha} = E_{B \alpha}+E_{C \alpha}$
(for the sake of simplicity, we omit the identity operators),
we obtain
\begin{equation}
f_{A}(\{E_{B \alpha}+E_{C \alpha}\}) = f_{B}(\{E_{B \alpha}\})
f_{C}(\{E_{C \alpha}\}) .   
\label{feq}
\end{equation}
Using the same method as AFLP$^6$, we can easily 
prove that Equation (\ref{feq}) has the unique solution---the 
three functions $f_{X}$ ($X=A,B,C$) must be exponentials:
\begin{equation}
f_{X}(\{E_{X \alpha}\}) = f_{X}(0) \exp\left[ \sum_{\alpha>0} 
\tau_{\alpha} E_{X \alpha} \right] , 
\end{equation}
where $f_{X}(0)$ is a normalization factor.
Here $\tau_{\alpha}$ are complex parameters which are
the same for the three systems $A$, $B$, and $C$. 
For example, in the case of the SU(2) group, we find
\begin{equation}
f_{X}(J_{X +}) = (1+|\zeta|^{2})^{-j_{X}}\exp(\zeta J_{X +}) ,  
\end{equation}
where $X=A,B,C$ and $\zeta \in \mathbb{C}$.
The operator-valued function $f_{X}(\{E_{X \alpha}\})$ for each 
of the three systems ($A$, $B$, and $C$) is precisely the function 
that produces the generalized CS.
(The factor $f_{X}(0)$ is recognized as the normalization 
factor ${\cal N}$).
This completes the proof of uniqueness.
\vspace{1.6\baselineskip}

\noindent
\uppercase{8. Conclusions}
\vspace{0.8\baselineskip}

\noindent
As was shown, the generalized CS keep two most important 
classical attributes of the Glauber CS.
(1) The known fact: each point $\Omega$ of the phase space 
corresponds to the coherent state $|\Omega\rangle$. If the 
Hamiltonian is linear in group generators, the 
generalized CS evolve along 
a classical trajectory $\Omega(t)$ in the phase space.
(2) The new fact: For the semisimple Lie groups,
the generalized CS are the unique Bell states, i.e., the
pure quantum states preserving the fundamental classical
property of satisfying Bell's inequality upon splitting.
\vspace{1.6\baselineskip}

\noindent
\uppercase{Acknowledgements}
\vspace{0.8\baselineskip}

\noindent
This work was supported by the Fund for Promotion of Research
at the Technion and by the Technion VPR Fund---Promotion of 
Sponsored Research.
\vspace{1.6\baselineskip}

\noindent
\uppercase{References}

\begin{flushleft}
1. H. Ezawa, A. Mann, K. Nakamura, and M. Revzen, 
   {\em Ann. Phys. (N.Y.)} {\bf 209}, 216 (1991). \\
2. E. Schr\"{o}dinger, {\em Naturwissenschaften} {\bf 14}, 
   664 (1926). \\
3. R. J. Glauber, {\em Phys. Rev.} {\bf 131}, 2766 (1963). \\
4. W.-M. Zhang, D. H. Feng, and R. Gilmore, 
   {\em Rev. Mod. Phys.} {\bf 62}, 867 (1990). \\
5. A. M. Perelomov, {\em Generalized Coherent States and 
   Their Applications} (Springer, Berlin, 1986). \\
6. Y. Aharonov, D. Falkoff, E. Lerner, and 
   H. Pendleton, {\em Ann. Phys. (N.Y.)} {\bf 39}, 498 (1966). \\
7. J. S. Bell, {\em Physics} {\bf 1}, 195 (1964-1965). \\
8. J. F. Clauser, M. A. Horne, A. Shimony, and
   R. A. Holt, {\em Phys. Rev. Lett.} {\bf 23}, 880 (1969). \\
9. A. Peres, {\em Am. J. Phys.} {\bf 46}, 745 (1978). \\ 
10. N. Gisin, {\em Phys. Lett.} A {\bf 154}, 201 (1991). \\
11. A. Mann, M. Revzen, and W. Schleich, {\em Phys. Rev.} A
    {\bf 46}, 5363 (1992).
12. C. Brif, A. Mann, and M. Revzen, {\em Phys. Rev.} A {\bf 57},
    742 (1998).
\end{flushleft}

\end{document}